\documentclass[tighten, times, twocolumn]{aastex631} 
\usepackage{graphicx}
\usepackage{color}
\usepackage{lineno}

\newcommand{\etal}{\mbox{\rm{et al.}~~}}
\begin{document}
\title{Dark Matter Genesis}
\shorttitle{Dark Matter Genesis}
\author{Jeremy Mould}
\affil{Swinburne University}
\affil{ARC Centre of Excellence for Dark Matter Particle Physics}

\email{jmould@swin.edu.au}

 \received{Nov 3 2024}
 \revised{Dec 2, 2024}
 \accepted{Dec 30, 2024}

\begin{abstract}
Recent discoveries
	of  primordial black hole (PBH) candidates by means of high cadence microlensing open the way to a physical understanding of the formation of dark matter as a chapter in the thermal history of the Universe. Two complementary sites of PBH formation are considered, inflation and the early Universe at TeV  to MeV energies. In the latter case the Friedman equation, together with mass 
	measurements,  
	reveal the threshold energy, the mass spectrum and the likely end point of this epoch. Some of the many recent exoplanet detections may conceivably have been detections of PBHs.
When the Universe cools to MeV temperatures, larger mass PBH would form similarly, reaching the supermassive regime. The discovery of numerous supermassive black holes (SMBH) at high redshift with JWST fulfils this expectation.
We corroborate the idea that Planck mass relics could be an important component of dark matter, and find that 
these are formed by PBH with initial mass less than approximately 6 $\times$ 10$^{-16}$  M$_\odot$ and cosmic temperature above 10$^9$ GeV.  Although in some mass ranges PBH can only make up a modest fraction of $\Omega_{matter}$, it is possible that all astrophysical dark matter, as distinct from axions and WIMPs, is of PBH origin.
\end{abstract}

\section{Introduction} 
Gravitational microlensing was first to show
that at least 80\% of galaxies' mass is non-baryonic 
(Alcock \etal 1998, Silk 2000), and this has been confirmed and strengthened by observations  
of the microwave background.
Following the possible discovery by Niikura \etal (2019) of a primordial black hole (PBH) with a mass in the range 10$^{-11}$ to 10$^{-6}$ solar masses and others at the high end of this range 
, we consider how such objects may be formed and evolve.
PBHs have long been associated with the  initial period of inflation that drove exponential expansion of the universe (Byrnes \& Cole 2021). Here we propose that these seeds grew during the radiation dominated era $\S$3 , and we follow their evolution to the present day, 
through recombination, the surface of last scattering of 
the remaining photons, to galaxy formation ($\S4)$ to the properties of
survivors today ($\S5$).
  We raise the possibility in $\S$5 that some  exoplanet discoveries may have been misclassifications of PBH.  
Our conclusions in $\S$6 include a proposed test for the PBH fraction of dark matter  via rotation curve measures of M* galaxies at z = 1. Forthcoming new  facilities for microlensing will advance this probe of the early universe.

\section{Primordial Black Holes as Dark Matter}

PBHs meet the basic criteria for a good dark matter candidate. They are cold, non-baryonic and stable, but they are not totally dark,
radiating with a 
temperature inversely proportional to their mass, so long as they remain above the Planck mass. Zel'dovich and Novikov (1967) and Hawking (1971) found that PBHs could form from overdensities in the early Universe. If they form before nucleosynthesis, PBHs are non-baryonic.  Recent reviews are by Green (2024) and Carr \& Kuhnel (2022).

In the Carr \& Hawking (1974) collapse model, the PBH mass M can be related to the horizon mass at the time of its formation,
M = $\kappa M_H $, 
  where M$_H$ = $c^3$/(2GH) is the horizon mass, and $\kappa$ is the efficiency of collapse.
From the conservation of entropy in
the adiabatic cosmic expansion, Carr \etal (2021b) find M $\approx$ 10$^{15}$ (0.05/k$_{PBH})^2$ M$_\odot$, where k is wavenumber in Mpc$^{-1}$ and k = aH/c, with H = $\dot{a}/a$ and a the scale factor, H being close to constant to drive exponential expansion. Gouttenoire \& Volansky (2024) find solar mass PBH
in supercooled phase transitions at TeV temperatures.
There are many proposals for PBH formation during inflation covering a range in k as broad as the different potentials under consideration.
Here we shall be agnostic about this, as we suppose that  PBHs formed during inflation are simply the seeds for density peaks present during the following radiation dominated era. A timeline for this hybrid approach to PBH formation is shown by Ai, Heurtier \& Jung (2024, their Figure 2).

\section{Physical conditions at DMG}
The Friedman equation for a radiation dominated Universe, expressed in terms of the Schwarzschild radius of the PBH is r = ct/8, where t is the time since the Big Bang. 
This yields
a time of formation for a 10$^{-7}$ M$_\odot$ object of 10 femtosec. At this time the cosmic temperature was 11~TeV. 
Lu \etal (2023) have proposed models in this energy regime for 10$^{-5}$, 10$^{-9}$ and 10$^{-12}$ M$_\odot$ PBHs. They invoke beyond the standard model (BSM) physics as the trigger at these energies.

Karam \etal (2023) make models in which high amplitude curvature perturbations lead to the formation of PBHs. These cover  a range of wavenumbers k from 10$^{13}$ to 5 $\times$ 10$^6~Mpc^{-1}$ and PBH masses from 5 $\times$  10$^{-14}$ to 0.5 M$_\odot$. The abundance of PBH is a strong function of the peak amplitude, and for a wide range of the PBH masses we are concerned with here, it seems possible to have $\Omega_{PBH}$ approach unity by choosing the peak amplitude.
The shape of the mass spectrum is $\sim$ M$^4$e$^{-7M/2}$, but this may result from the choice of the Press-Schechter (1974) formalism. 

%
As an alternative to Press-Schechter (PS) we propose a geometric
formalism. PS asks what is the likely spectrum of density peaks ? Instead,
we ask, once a large spherical PBH has formed, how many smaller ones can be
fitted around it ? That said, we cannot neglect that the physics of these density peaks
is unclear. The QCD phase transition has been advanced as a possible site (e.g. Musco, Jedamzik \& Young 2024; Carr \etal 2021a), as
the short range nuclear strong force suddenly yields to the cosmic expansion. In addition
to this problem, there is the question of how PBH of any mass have avoided leaving
a signature on Big Bang Nucleosynthesis, noted first by Bicknell \& Henriksen (1979).

\subsection{PBH initial mass function}
Assuming that PBHs actually are a significant fraction of galaxies' dark matter, packing theory\footnote{https://www.vedantu.com/chemistry/unit-cell-packing-efficiency} can be applied, with PBHs forming in random parts of the volume that we nominally call overdensities, although we do not track density fluctuations in this toy model, the only condition being that they do not overlap. 
Packing theory quantifies the notion that in a given space more small objects
will fit than large objects as, N = 1 - ln f where f is the fractional area/volume/radius, independent of dimensionality\footnote{ 
For example,  for area, dn = (A$_0$ - A(f))/fA$_0$ = -df/f.
Apply that to two values of the scale factor a, a+da:  
n(a) =1 -ln f(a) and  n(a+da)=1 - lnf(a+da). So
dn(a)=ln (f(a+da)/f(a)) = ln (da).
Integrate from a0 to a: n(a)-n(a$_0$)= a ln a -a -a$_0 ln a_0 + a_0$. 
Suppose a = a$_0$x; a$_0$ = a/x; then N = n(a)-n(a$_0$) = ln x -a.
For N $>>$ ln x, log N = -log a + constant.},
and with f $<$ 1, then N $>$ 1
For scale factor a $\sim~\surd$t in the radiation era, and a $\sim e^t -1 \approx$ t for small t in inflation. 
In the latter case log N +log t = constant and log N + log M = constant.
For a $\sim~\surd$t in the radiation era (RE), 2 log N + log M = constant.
Both power law IMFs are shown in Figure 1. Measuring the mass function of
PBHs can tell us whether a PBH burst was seeded during inflation or wholly the
progeny of the RE. Carr \& Silk (2016) recommend a power law IMF for PBH.

The packing process stops when it is impossible to fit any more of the steadily larger PBHs demanded by r $\sim$ t in the co-moving volume. Figure 1 shows the (log N, log M) mass distribution that results. The masses have been normalized so that the most likely mass is equated to the 
high end of the Subaru team mass range. 
This identifies the time scale of the formation process and the temperature of the Universe during that time. Without it, there is nothing in the toy model but dimensionless radii, which are proportional to the mass, and thus normalizable to it.

The peak and rapid fall of the mass distribution means that it was most likely that the first PBH microlens to be detected would be close to the maximum mass to form,   
 even though the new high cadence experiment, in contrast to earlier work (Alcock \etal 1993; Paczynski \etal 1996), was sensitive to lower masses. This makes normalization to the very first PBH to be announced as such less hazardous than if the (log N, log M) distribution were flat. 

Since r $\sim$ M $\sim$ t, once the formation process has started, progressively larger PBHs are formed, steadily filling the Universe. The matter that is not subsumed into PBHs is free to move on to the next stages in the thermal history of the Universe, baryogenesis and primordial nucleosynthesis, which occurs one second later.
In what follows we consider PBH dark matter to be a significant part of galaxy halo dark matter. It is not necessary that it be dominant, and it can coexist with non-baryonic particle dark matter.  
The fraction of dark matter in the mass range we discuss that is PBH can be determined by high cadence microlensing experiments.
We emphasize that both PS and packing theory are formalisms
rather than strict physical theories. We see them as alternatives
and even complementary. Mould \& Batten (2024) show that the packing theory
IMF leads to an excellent fit of the QSO luminosity function, if the
supermassive black hole (SMBH) nuclei are large PBH.

\subsection{Other formation theories}
Carr \etal (2019) favor a multi-modal PBH mass spectrum with peaks at 10$^{-6}$
, 1, 30, and 10$^6$ M$_\odot$.
This suggests a unified PBH scenario which naturally explains the dark matter and recent microlensing observations, 
and the origin of the SMBH in galactic nuclei at high redshift.
By comparison the models of Lu \etal are also peaked at the energies where they assume a phase transition, except for the lowest mass model (10$^{-12}$ M$_\odot$) which is broader and rises to lower masses. They draw on the PS 
formalism, and find the PBH fraction of the dark matter to be 100\% in their lowest mass case, 0.5\% in the 10$^{-9}$ case and 2\% in the highest mass case. The latter two cases would be hard to detect in microlensing experiments, as the event rate towards the LMC $\sim$ 1.6 $\times$ 10$^{-6} \surd(M_\odot /M)$ per year for a 100\% PBH Galactic halo (Moniez 2001). The lowest mass is a problem observationally because of finite source effects. That is to say, the Einstein radius has become so small that an amplification event has begun to turn into a transit.
An alternative to PS is peak theory in which the number density is calculated of 
peaks where compact objects like galaxies and PBHs are expected to form, 
if the density contrast  exceeds some threshold (Wang \etal 2021). 
With 10$^{-13}$ and 30 M$_\odot$ models they achieve 2 or 3 orders of magnitude higher PBH production, 

\begin{figure}	
\includegraphics[width= .6\textwidth,angle=-0]{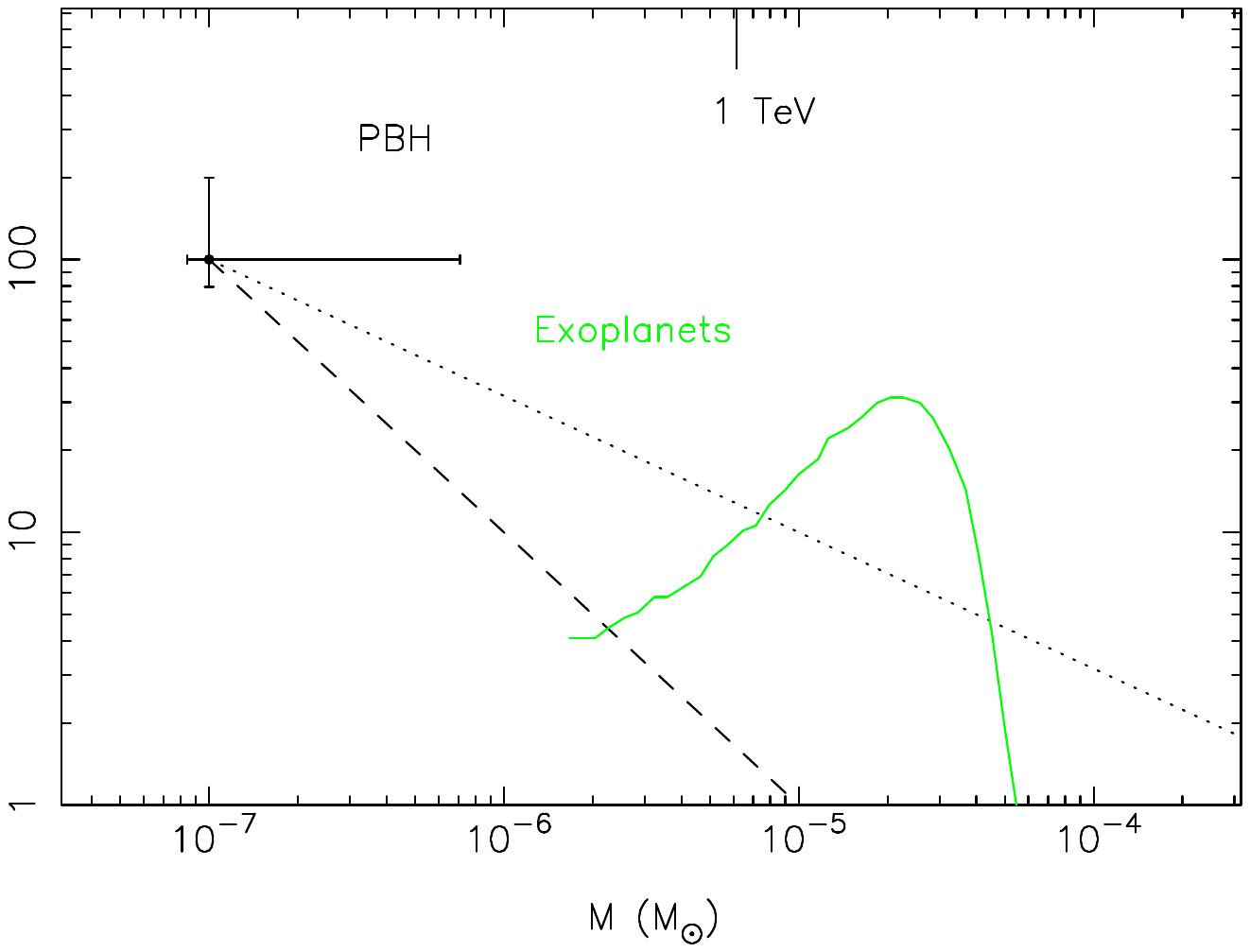}
\caption{On the low mass side the mass distribution (IMF) is of the form $\log N + \log M$ = constant, illustrated by the dashed line. There is a sharp cutoff on the low mass side where the PBH with the largest Schwarzschild radius is the first mass to be accommodated. A second IMF 2$\log N + \log M$ = constant is the dotted line.
	The limitations of this toy model are outlined in $\S$3.3. The point at 10$^{-7}$ M$_\odot$ is close to an IMF peak 
	proposed by Carr \etal (2019). The green dashed curve is a $Kepler$ exoplanet mass function from Suzuki \etal (2016).
In $\S5.2$ we ask whether any PBH microlensing events have been misclassified as exoplanets. If this happened to be the shape of the curve of a PBH mass function, it could be the signature of a burst of PBH formation later than that shown. 
 As the mass axis is also a cosmic cooling axis, an energy of a TeV 
 is marked.}  
\end{figure}
\subsection{Lessons from the toy model}
What packing theory fails  to predict accurately is the fraction of mass in the Universe that is PBH dark matter. Assuming random relative positioning of PBHs, the packing fraction is expected to be 64\%, whereas what is observed for the sum of all species of non-baryonic dark matter is 81\% (Ade \etal 2020). 
A physical model would trigger PBH formation at
density inhomogeneities and could make a successful prediction of a high dark matter fraction.
Nevertheless, the geometric toy model motivates the investigation of that physics and strongly suggests that it is the finite radius and stability of black holes that determines how many baryons remain in the  Universe to go on to make nuclei, atoms, galaxies and us. Everything we see around us may be just the leftovers from the primordial black holes which attempted to consume most of the Universe in its first seconds, were involved in seeding SMBH  during primordial nucleosynthesis, hosted  and controlled\footnote{By "controlled" we mean that mass losing PBH can provide the feedback mechanism that calls time on galaxy collapse.} galaxy formation when cosmic temperatures had cooled sufficiently, and are still around maintaining galaxies today.

To sum up, the Friedman equation for the radiation dominated era is prescriptive, but insufficient. 
At any time t in this energy range there is a specific radius of PBH that $can
$ be formed, r~=~ct/8. Whether a PBH $does$ form at that time involves TeV physics, possible seeding during inflation, and density inhomogeneity, as well as geometry.
\section{PBH evolution}
PBH evaporate with age as kT = $\hbar c / 4\pi r$ (Hawking 1971).    
Evolutionary tracks are shown in Figure 2 for a number of different masses. 
As the universe ages according to the Friedman equation through its radiation, matter and dark energy dominated stages, the PBH sticks to its Hawking equation. We used the mass loss rates of Mosbech \& Picker (2022) to evolve ten different masses between 10$^{-23}$ and 10$^{-1}$ M$_\odot$.  A dimensionless mass loss rate dlogM/dlogt is more suitable to our dlogkT time steps. For reference, dlogM/dlogt $\approx$ -1 for main sequence stars, their luminosity being c$^2$ dM/dt, where dM is the mass loss due to fusion. Intermediate mass PBH have dlogM/dlogt = --13.16 $\alpha_{24}$ (m20)$^{-3} ~(kT)^{-2}$, where kT is the cosmic temperature in eV (the x-axis in Figure 2) and m20 is the mass in 10$^{-20}$ M$_\odot$ units. 
For log M / M$_\odot~ > $ -15.3, $\alpha_{24}$ = 1, and for smaller PBH Mosbech \& Picker give a fit to it. While the above relation is strictly for RE, similar expressions can be written for the matter and dark energy dominated universe.

The lowest mass PBH do not reach matter/radiation equality before evaporating off the plot.
\subsection{ Recombination}
For any temperature of the Universe
there is an mn PBH that is evaporating at that time. The track becomes almost vertical at that kT. And this is true for kT = 13.6 eV for a PBH mass of m20. The duration of recombination is approximately from 13.6/kT-1 to 13.6/kT+1. That neglects electron density n$_e$ and T$^{3/2}$ which shorten the time a bit, as n$_e$ decreases as the scale factor a increases. The extent of $\delta$T is dlnT =2k/13.6= dln(1+z)=dlnt /2 for the RE. It is easy to show\footnote{
For PBH initial mass M0 and a constant k,
dM/dt = -k/M$^2$ and
dlmdlt=dlogM /dlogt = -kt/M$^3$.
Integrate: dM M$^2$ =-kdt, to get (M0$^3$-M$^3$)=3kt.
Divide: 3dlmdlt=(M0/M)$^3$-1.
So dlmdlt starts at zero, and at its half mass time is 7/3.   It continues to rise.....
} dlogM/dlogt = 7/3 at the half mass point of a PBH. So the mass range evaporating during recombination is $\delta$ln M = 17.2/13.6~14/3~10$^{-5}$. 

PBH ionise their environment when they evaporate to x=0.19, the baryonic mass fraction, where x is the degree of ionisation. So put 13.6/$\delta$(kT) = 0.19, the temperature change required to compensate for PBH ionisation. That is 13.6 dlnkT = 0.19 or dlogkT = 0.006, barely visible on our tracks. Or $\delta\log$t $\approx$ 0.01. And that is also the change in the age of the Universe and H$_0$, 2.3\%. An exact treatment of recombination with PBH evaporation is warranted
{Seager, Sasselov \& Scott 1999, Slatyer 2015, Batten \& Mould 2024).}

This delay of recombination is analogous to some of the varied solutions to the Hubble tension considered by di Valentino \& Silk (2023).
Although PBH are mentioned in their review, the ionization time bomb mechanism described here is different.  Scars that might be left on the surface of last scattering also need examination.

\subsection{Matter domination}
The next transition for larger PBH is the end of the RE at z $\approx$ 3400 in the Planck cosmology (Ade \etal 2020), which we use throughout. 
The second is crossing the surface of last scattering of the cosmic microwave background (CMB). Next is redshift 100, close to the free fall time for galaxies, which moves left or right depending weakly on the mass under consideration. Low mass PBH reach that line (together with accompanying WIMPs) to form dark matter halos into which the baryons will fall.
Finally, the start of the dark energy epoch is defined by a$_{DE}^3 = \Omega_m/\Omega_\Lambda$ = 0.43. 
From the start of matter domination (MD) to the onset of 

We adopt a nomenclature for PBH in which m11 is a PBH of mass 10$^{-11}$
M$_\odot$ and M7 has 10$^7$ M$_\odot$.
We note that PBH smaller than m16.8 have now reached Planck mass, m$_P$ = 22 $\mu$g. Profumo (2024) discusses whether PBH stall at m$_P$ or continue on to lower masses\footnote{An h$\nu$ = kT photon at the Hawking temperature has an energy of 200 m$_P$, seemingly a hard task for a Planck mass particle to emit.}. 
If PeV physics and beyond permits them to be made, they could number up to 10$^{50}$ in the Galaxy (see also MacGibbon 1987 and Taylor \etal 2024). At 
10$^{-20}$ the radius of the proton, these objects would pass through ordinary matter, if geometric cross sections are a guide.
Planck mass relics have an area 
of 10$^{-65}$ cm$^2$, and the reaction rate with gas per unit volume 
would be below 
10$^{-10}$ per year.
Such relics in the Milky Way might contribute to the cosmic ray rate, but at 22$\mu$g are ultrarare and 10$^{10} $ times more energetic than the OMG CR observed in 2021. 
As non-baryons their interaction with normal matter is outside the scope of this paper, but in
a sense they might be regarded as fundamental particles with a macroscopic mass, and 
could even be described as ultramassive WIMPs.

The dependence of the morphology of Figure 2 on fundamental constants is interesting.
The Boltzmann constant zeropoints both axes equally.  The Planck constant controls the rate of cooling of the PBH.
Newton's constant controls the time/temperature x-axis.
To the left of the radiation to matter transition that is all there is.
$\Omega_ {matter}$ affects the slope thereafter, but $\S$3.1 
 hypothesises that $\Omega_{ matter}$ is determined by something as simple as packing geometry. 

\begin{figure*}	
\includegraphics[width=1.2\textwidth,angle=-0]{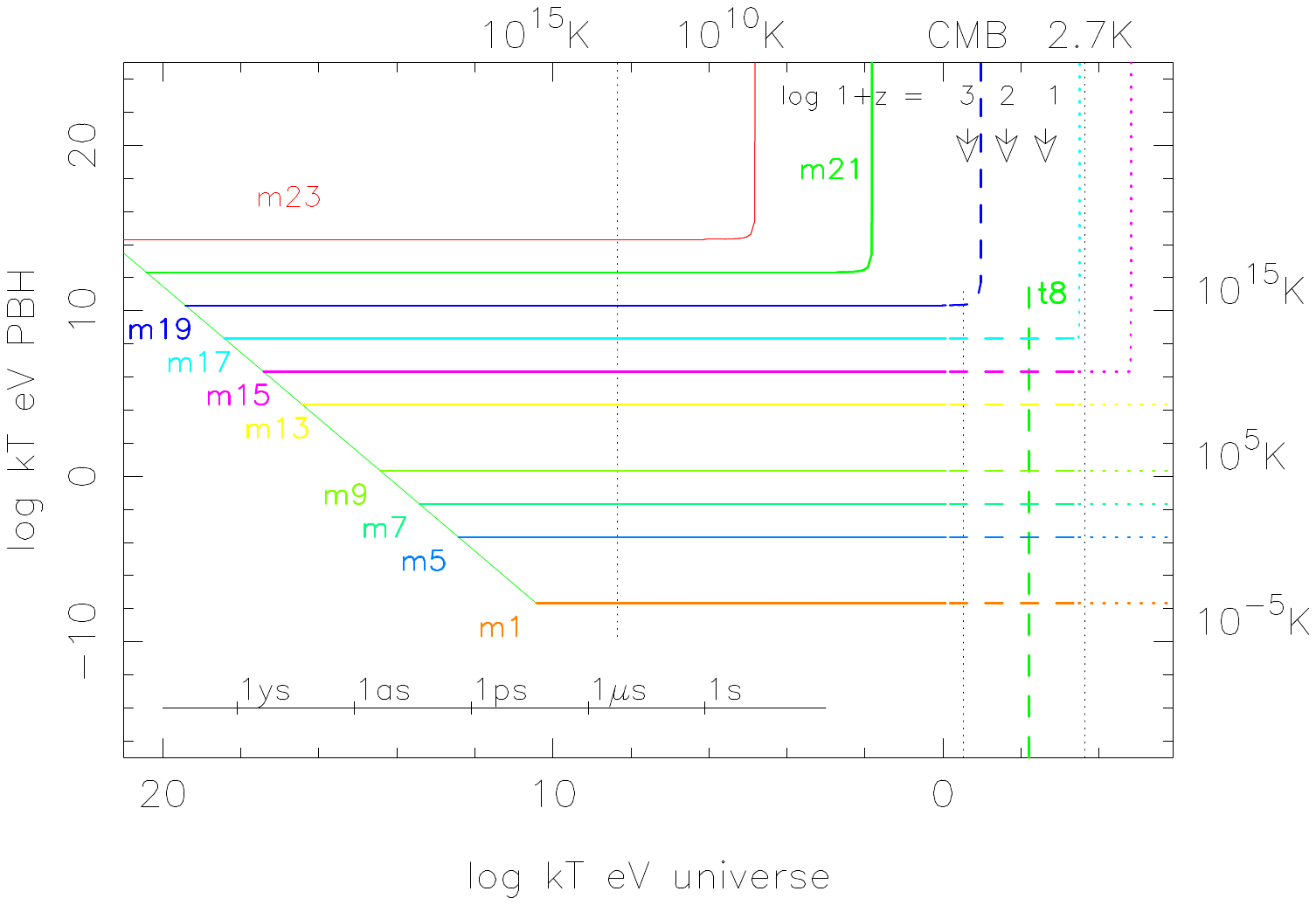}
\caption{Evolution of 10 different mass PBH in PBH temperature vs cosmic temperature.
Kelvin units are on the top and right borders.
PBH changing their evolutionary rate as T $\sim$ M$^{-1}$, from radiation (solid line), to matter (dashed line), to dark energy dominated (dotted line). 
	The t8 line marks the start of galaxy formation. 
  The three dotted vertical lines are from left to right the QCD transition at 220 MeV, the surface of last scattering of the CMB 
  and the temperature today, 2.7K. The first of these is discussed as a possible PBH trigger by Alonso-Monsalve \& Kaiser (2023). 
As the universe cools, the lowest mass PBH take a right angle turn to high temperatures, rapidly becoming Planck mass relics.}
\end{figure*}
\subsection{Galaxy formation}
From the end of RE at t$_{eq}$ 
there is an infall radius of matter for a PBH of mass M of g$t^2$/2, with gt $<<$ c, where  $t$ is the time since the end of RE and g is the gravitational acceleration into the PBH, and (until galaxy formation) a corresponding infall. 
These infall regions can attain kpc size in Myrs, making them protogalaxy candidates. Closer to the PBH itself,
a photosphere would attain hydrostatic equilibrium with pressure P = g m$_e/\sigma_e$, where these quantities are the mass and Thomson cross section of the electron. P =1.5 $\times$ 10$^{-4}$g and g = c$^4$/GM = 6 $\times~10^{17}$/M cm s$^{-2}$ 
with M in solar units. 
This would slow the timescale for mass loss to a diffusion time, as is the case for conventional stars.  Radiative transfer in a relativistic infalling gas is beyond the scope of this paper, but for masses larger than the subsolar range  in figure 2, is important to explore in the context of proto-galactic nuclei.
After galaxy formation the density of matter surrounding a PBH drops to that of the intergalactic medium (IGM), which, for $\rho_{IGM}~ << ~ \rho_{galaxies}$, is insufficient to sustain infall. 

Like other forms of dark matter, at $\sim $ 10$^8$ years PBH collapse into a potential well of their own making. PBH that lose mass in the galaxy formation period (30, 3) in redshift have a small range in mass between m$_1~\sim$ m17 and m$_2~\sim$ m19.
For PBH the mass loss is dlogM/dlogt =
$\sim$ --(kT$_{PBH}$/10 GeV)$^3$, a runaway rate. 
We carried out a test with a 100,000 particle scale free dark-matter-only n-body simulation. 
Similar structures resulted in both the mass losing and the mass conserving case.  
A 3\% increase in the circular velocity of the resultant potential resulted from a 12\% loss in total mass.
Effectively, mass losing PBH are providing feedback, often attributed to mechanisms not intrinsic to the collapse process.

\subsection{Matter density stability}
If PBH are a significant fraction of the dark matter and they lose mass, how come $\Omega_m$ measured in the cosmic microwave background (CMB) $~\approx~\Omega_m$ now ?
The answer may be in the mass range that has lost mass between passing
the redshift of the surface of last scattering and the current time. The range is m15.2 to m19.2, and those PBH would have lost essentially 100\% of their mass in that period. We have no observational data on that mass range, only broadly on the range m11 to m6 which are too large to evaporate, as dM/dt $\sim $ M$^{-2}$. 
 In the range that has lost mass since the CMB redshift the limit on $\Omega_m$ is an integral over the PBH mass range, m1 to m2
 $$\frac{\Delta\Omega_m }{\Omega_m } = 1.75 f \int_{m1}^{m2} dm~~\frac{~2~{\rm ln10  } ~dm}{ m1+m2} 
 = 1.75 f~{\rm ln10 }~\frac{m2-m1}{m1+m2}$$
 $$= 1.75 f~{\rm ln 10 }~
 ~$$
 The uncertainty in $\Omega_m$ found by Davis \etal (2024) in the Dark Energy Survey then requires 1.75 f ln 10 $<$ 0.017/ 0.27 and f $\lessapprox$ 1.5\% 
 for a flat, constant equation of state cosmology. The limit becomes 7.5\% in a variable w (equation of state) cosmology.
 This can be relaxed (Efstathiou 2024) by allowing a 3\% variation of the SNIa standard candle\footnote{https://www.issibern.ch/teams/shot/wp-content/uploads/sites/186/2024/07/issi.pdf}. The conserved majority share of $\Omega_m$ would be a combination of PBH outside the m15 to m19 mass range, WIMPs and axions.
Micro-PBH, which reach Planck mass before they come to the CMB line in Figure 2 may be an important contributor with a maximum initial mass  $\approx $ 8 $\times$ 10$^{-20}$ M$_\odot$.
\subsection{The bandwidth of the PBH IMF}
Finite source effects reduce the optical depth in the LMC line of sight for low mass PBH. A further, but speculative, reduction may come about from
the PBH IMF. If every decade in mass has equal mass (n $\sim$ m$^{-1}$), and
a given microlensing experiment is sensitive to 2 decades in cadence, then only 4 decades in mass from the range m17 to M3 are covered, and the maximum PBH fraction of the dark matter, f, visible in that experiment is 20\%. We chose m17, as that is the smallest mass that has not already fallen to the Planck mass. 
If PBH formation were to begin at the start of the RE at log kT $\gtrsim$ 21, f $<$ 10\% for an individual experiment. 
If the PBH IMF consists of discrete pieces, triggered by events in the RE, this limitation does not apply.
\section{PBH today}
The PBH IMF of Carr \etal (2019, their figure 2) shows one peak
in the region of m6 PBH. These PBH should be detectable now.
\subsection{The HR diagram for compact objects}
Luminosities for PBHs were calculated  from their radii and temperatures 
at the current epoch and are shown in Figure 3 together with the nearest white dwarfs and magnetars plus pulsars from the fundamental plane of Chu \& Chang (2023). To eliminate the third parameter, we chose a typical 1.5 km radius. The m5 --m13 labels denote the initial and current masses of these objects. A close by m5 or m3 object would in principle be observable, 
The extraordinarily high temperatures predicted for PBH of lower mass than Figure 3 accommodates might not be possible for baryonic stars due to neutrino cooling.
The bolometric correction for optical observers, moreover, for a very hot m13 PBH 
is  of order 12.5 magnitudes. 
A further possible observational quirk of low mass PBH is that as they approach 10$^{17}$ m$_P$ (1.5 Mton), their radiation may become noticeably quantized to $\sim$1 photon/second.
\begin{figure}	
\includegraphics[width=1.2\textwidth,angle=-0]{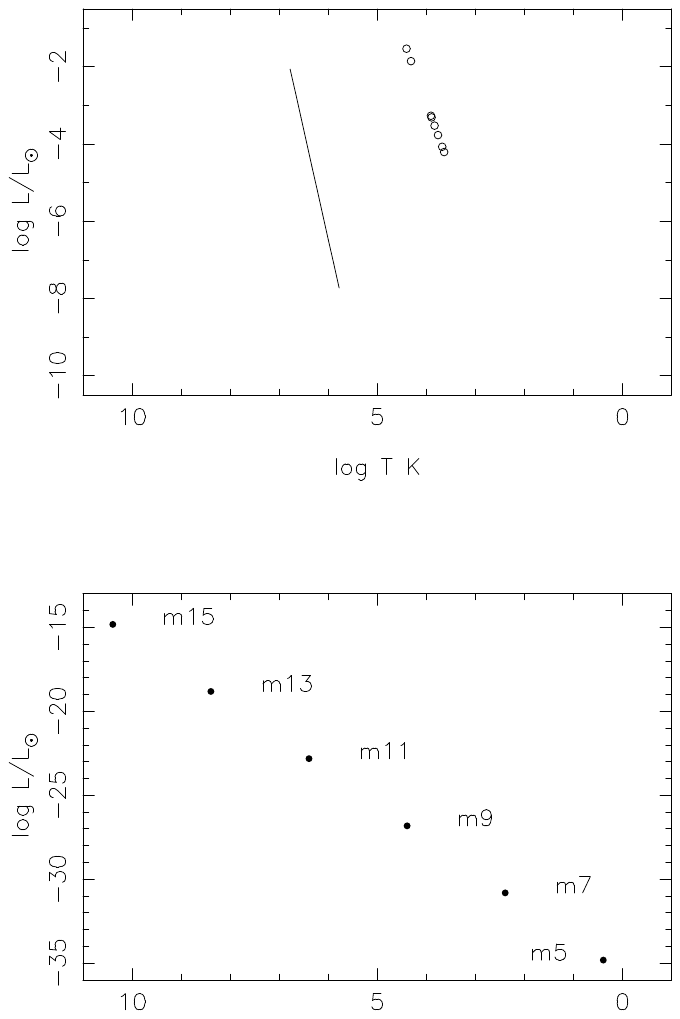}
\caption{The Hertzsprung Russell diagram for PBHs of observable mass. Also shown is the nearby white dwarf cooling sequence (open circles) and pulsars plus magnetars from a linear fit by Chu \& Chang (2023). 
	The baryonic stellar remnants are cooling sequences. 
 An m7
	has a brown dwarf temperature. Pressure broadening at log g = 36 would erase all spectral lines, however. 
 PBH are the most compact of compact objects.}
\end{figure}
There are objects in the Gaia DR3 archive at 1--10 $\times$
 10$^{-6}$ L/L$_\odot$, but candidates will
require spectroscopic sorting from brown dwarfs, which are surprisingly blue at Gaia wavelengths.

\subsection{Have PBH been detected towards the Galactic center ?}
 Some of the 200 exoplanet detections (Mroz \& Poleski 2024) in the line of sight to the Galactic bulge may have been PBHs, as the dark matter density peaks at the Galactic center. Based on optical depth calculations that fraction is approximately 30 $\pm$ 5\% for a fully PBH halo and 3\% for a 10\% PBH halo. We discount possible  Planck mass relics in this estimate. To calculate
 baryonic and non-baryonic $\tau$, we used the 
 Galaxy model by de Jong \etal (2010) 
 but, to err on the side of caution, put a core on its NFW dark matter component, which otherwise has a cusp at the center of the Galaxy. Such cusps are not seen to last in galaxy evolution simulations. Adding a bulge from Binney \& Vasiliev (2023), while supplying the source stars, made a very small change to the baryonic optical depth, because of its steep fall along the Baade's Window line of sight to the observer, which was used for the calculations. Uncertainty about the behavior of the scale height of the disk  ($\pm$ 50 pc) at Galactic radii interior to the Sun's imposes the nominal limits we have attached.

Figure 1 also shows a $Kepler$ transit exoplanet mass function by Suzuki \etal (2016). 
Their M type primaries function was chosen, as these are the commonest stars in the Galaxy, but the G type function is similar with a somewhat higher peak mass and lower frequency of exoplanets per star. 
Both are consistent with microlensing exoplanet observations. We do not know if the early Universe  made 10$^{-5}$ M$_\odot$ PBH, but if it did, their distribution function could fit within that curve. High cadence LMC microlensing 
would not have detected them, because the event rate from 10$^{-5}$ M$_\odot$ PBH would be a tenth that of the rate at 10$^{-7}$ M$_\odot$ (Moniez 2021). 
The combined Magellanic Cloud microlensing campaigns would have detected Jupiter mass PBHs, if they were present (Alcock \etal 1998 ; Alfonso \etal 2003).
Carr \& Kuhnel (2022) and Alonso-Monsalve \& Kaiser (2023) proposed the QCD phase transition at 220 MeV as a trigger  for PBH formation. This would have produced 0.16 M$_\odot$ M dwarfs, also not seen in LMC microlensing.
The next significant energy level which the Universe passes through is 0.511 MeV. This corresponds to 1.4 $\times$ 10$^7$ M$_\odot$ PBH, or Supermassive PBH (Bicknell \& Hendriksen 1979; Soltan 1982; Mould \& Batten 2024), and Sobrinho \& Augusto (2024) discuss this as an accompaniment to primordial nucleosynthesis. It is a possible path for solution of the early SMBH problem identified by JWST (Labb\'e \etal (2024).

Sumi \etal (2023) fitted a Free Floating Planet (FFP) initial mass function to the results of the MOA 9 year microlensing survey (Koshimoto \etal 2023). The shortest Einstein time found in this survey was 0.057 days, twice the 0.0288 days for the 10$^{-7}$ M$_\odot$ Phoebe. The Einstein time $E_T~\sim~\surd M$, where M is the mass of the lens. Acceptable fits rise sharply at masses between 3 $\times$ 10$^{-5}$ and  
 10$^{-7}$ M$_\odot$, depending on the parameters of the mass function. Microlensing FFPs can only be loosely defined as free floating, as they are required to show no underlying amplification of a host star with longer t$_E$ in the light curve, a condition which can be found from simple analytical considerations to be satisfied by placing the planet beyond a Mars orbit for a 1 M$_\odot$ host. Be that as it may, PBHs may masquerade as FFPs in their light curves, and the Sumi \etal mass distribution function may be occupied in part by PBHs at the level we have estimated from Baade's Window optical depths.

 \section{Conclusions}
Setting aside the ambiguity of microlensing events from the Galactic bulge, \begin{itemize} \item the possible discovery of PBHs in 
			the mass range 10$^{-11}$ to 10$^{6}$ M$_\odot$ places the formation of some of the dark matter as 
			early as 
			the quark epoch before baryogenesis at the electroweak scale (100 GeV), between the end of inflation and attoseconds 
 after the Big Bang 
			and as late as the primordial nucleosynthesis era. %
		\item 
			 PBHs below 10$^{-17}$ M$_\odot$ would now be Planck mass relics and could account for a significant fraction of today's dark matter. These may have been seeded in a still earlier inflation scenario which seems capable of a wide range of formation predictions, both in mass and abundance 
			 \footnote{
 We could foresee in this scenario a particular PBH mass spectrum might point to a feature in the potential. Peaks in the power spectrum can be induced by a bump, dip or step in the potential (Karam \etal 2023).
		}
	\item Some of the exoplanet detections in the line of sight to the Galactic bulge may have been PBHs, as the dark matter density peaks at the Galactic center. 
	\item	Annihilation radiation from the Galactic center may be an associated dark matter signal (Song \etal 2024).  \item Further observations 
 to populate the mass spectrum and simulations that incorporate the physics of this energy domain, with geometrical distributions of PBHs other than uniform, are desirable to explore this chapter of the thermal history of the Universe fully. A sharp cuton could be evidence for the role of geometry as outlined here. 
		\item
 Results from supernova cosmology (0.1 $<$ z $<$ 1 ) and the CMB regarding the critical mass fraction in matter ($\Omega_m$)
 suggest that only a modest percentage of the overall dark matter can be in the form of mass losing PBH with masses between m15 and m19. \item The PBH era could seed the supermassive black holes that seem to be numerous at high redshift, if it extends to times of a few seconds (Mould \& Batten 2024). \item  Mass loss by m20 PBH could delay recombination, affecting the CMB H$_0$.

 \item Dark matter halos with a fraction f of PBH  in the range m15.3 to m17.3 may have lost that fraction of their mass  to evaporation  (to Planck mass relics)
	in the last 12.6 Gyr (i.e. since z = 1). 
	Their v$^2$r should have been 100/f \% larger than they are today, where v and r are the velocity and radius at the flat part of their rotation curves.
	Twenty one  cm line width
	measurements of the Schechter function's M* galaxies could test this prediction.
\item Observations of the low mass end of the PBH spectrum below 10$^{-8}$ M$_\odot$, which can tell us more about the early Universe, are more difficult for  Magellanic Cloud microlensing fields because of finite source star effects, which reduce the optical depth to lensing. 
But there is telescope construction underway that may meet the challenge (DeRocco \etal 2024).
\item      
	The cold dark matter paradigm is a minimalist theory, successful over  almost fifty years. Additions are
now being trialed, and the rich properties of PBH make them good candidates, even
though their formation physics remains to be fully understood.

\end{itemize}
\section*{References}
\noindent Ade, P. \etal , Planck collaboration 2020, A\&A, 641, A6\\
Ai, W-Y, Heurtier, L \& Jung, T-H 2024, arxiv 2409.02175\\
Alcock, C. \etal 1993, Nature, 365, 621\\
Alcock, C. \etal 1998, ApJL, 499, L9\\
Alfonso, C. \etal 2003, A\&A, 400, 951\\
Alonso-Monsalve, E. \& Kaiser, D. 2023, Phys Rev L, 132.231402\\
Batten, A. \& Mould, J. 2024, in preparation\\
Bicknell, G. \& Henriksen, R. 1979, ApJ, 232, 670\\
Binney, J. \& Vasiliev, E. 2023, MNRAS, 520, 1832\\
Byrnes, C. \& Cole, P. 2021, arXiv 211205716\\
Carr, B. \etal 2021a, PDU, 31, 00755\\
Carr, B. \etal 2021b, RPPh, 84, 6902\\ 
Carr, B. \etal 2019, Physics of the Dark Universe,31, 755\\ 
Carr, B. \& Hawking, S. 1974, MNRAS, 168 399\\
Carr, B. \& Kuhnel, F. 2021, arxiv 21100282\\
Carr, B. \& Silk, J. 2016, MNRAS, 478, 3756\\
Chu, C-Y \& Chang,H-K 2023, MNRAS, 526, 1287\\
Davis, T. \etal 2024, ApJ Letters in press\\
de Jong, J. \etal 2010, ApJ, 714, 663\\
DeRocco, W. \etal 2024, PhRvD.109b3013\\
Efstathiou, G. 2024, arxiv 2408.07175\\
Green, A. 2024, Nuclear Physics B, 1003, id.116494\\
Gouttenoir, Y. \& Volansky, T. 2024, arxiv 2305.04942\\
Hawking, S., 1971,  MNRAS, 152, 75\\
Labb\'e , I. \etal 2024, Nature, 616, 266\\
Koshimoto, N. \etal 2023, AJ, 166, 107\\
Karam, A. \etal 2023, JCAP, 3, 13\\
Lu, P. \etal 2023, Phys Rev Lett, 130, 1002\\
MacGibbon, J. 1987, Nature, 329, 308\\
Moniez, M. 2001, Proc. XXXVth Rencontres de Moriond, Tran, Mellier \& Moniez. Les Ulis: EDP Sciences\\
Mosbech, M \& Picker, Z. 2022, SciPost, 13, 100\\
Mould, J. \& Batten, A. 2024, submitted to ApJL\\
Mroz, P. \& Poleski, R. 2024,  Handbook of Exoplanets, 2nd Edition, Deeg and  Belmonte (Eds), Springer\\
Musco, I., Jedamzik, K. \& Young, S. 2024, Phys Rev D. 109, 3506\\
Niikura, H. \etal 2019, Nature Astronomy, 3, 524\\
Paczynski, B. \etal 1996, IAU Symposium, 169, 93\\
Press, W. \& Schechter, P. 1974, 187, 525\\
Profumo, S. 2024, arxiv 2405.00561\\
Rezazadeh, K. \etal 2022, EPJ C , 82, 8, id.758\\
Seager, S., Sasselov, D. \& Scott, M. 1999, ApJ, 523, L1\\
Silk, J. 2000,
Proc. 7th International Symposium on Particles, Strings and Cosmology,  Eds: K. Cheung, \& J. Gunion\\
Slatyer, T. 2015, arxiv 1506.03812\\ 
Sobrinho, J. \& Augusto, P. 2024, MNRAS, 531, L40\\
Soltan, A. 1982, MNRAS, 200, 115\\
Song, D. \etal 2024, MNRAS, 530, 4395\\
Sumi, T. \etal 2023, AJ, 166, 108\\
Suzuki, D. \etal 2016, ApJ, 833, 145\\
Taylor, Q. \etal 2024, PhysRevD, 109, 406\\ 
Wang, Q. \etal 2021, Phys Rev D, 104, 8, id.083546\\
Zel'dovich, Y.B. \& Novikov, I.D. 1967, Sov. Astron. 10, 602\\
Zheng, R. \etal 2022, Chinese Physics C, 46, 4, id.045103\\

\subsection*{Acknowledgements}
The ARC Centre of Excellence for Dark Matter Particle Physics is funded by the Australian Research Council. Grant CE200100008. Simulations were carried out on 
Swinburne University's Ozstar \& Ngarrgu Tindebeek supercomputers, the latter named by Wurundjeri elders and translating as "Knowledge of the Void" in the local Woiwurrung language. I thank  colleagues in our microlensing team and the CDMPP for helpful discussions and the referee for improvements to the paper.
\subsection*{Code availability}
The evolutionary tracks and toy model with README files
 are available at https://github.com/jrmould
  and 10.5281/zenodo.14195697
\section*{Appendix}
The two equations assumed in $\S$2 (and why r = GM/c$^2$  $\sim$ t) are the Friedman equation for the radiation dominated era, giving the energy density, to be multiplied by Hubble volume.
$$\frac{3}{32\pi Gt^2} (tc)^3 4\pi/3 = {\rm M(PBH)}$$ and
$$t \approx ~1.3~\left (\frac{1 MeV}{kT}\right )^2 ~{\rm s}~~
 = 1.3\left (\frac{1 GeV}{kT}\right )^2 ~{\rm{\mu s}}~~~
 = ~1.3\left (\frac{1 TeV}{kT}\right )^2 ~{\rm ps}$$
from radiation energy density = 4$\sigma T^4/c$ and the Friedman equation. 
Similarly for PeV and attoseconds.
In inflation we have from Carr \& Hawking
$$M/M_\odot = 1.13 \times 10^{15} ~~~\frac{\kappa}{ 0.2} \left(\frac{g*}{106.75}\right )^{-1/6} \left(\frac{ k*}{k_{PBH}}\right )^2$$ where $\kappa$ is efficiency of collapse, g* is number of relativistic degrees of freedom and k$_{PBH}$ is the wavenumber of the PBH as it exits the horizon and k* = 0.05 Mpc$^{-1}$. 

The relationship between PBH initial mass and energy is M/M$_\odot$ = 3.3 $\times$ 10$^{-6}$ (1~TeV/kT)$^2$.
The PBH range 5 $\times$ 10$^{-7}$ to 
 5 $\times$ 10$^{-9}$ M$_\odot$ is often called the PBH asteroid range (e.g. Caplan \etal 2023). 
In the temperature range, 100 down to 1 TeV, the Universe passed through the PBH initial mass range
 3 $\times$ 10$^{-10}$ to  
 3 $\times$ 10$^{-6}$ M$_\odot$.
Three cooling channels are available to it, expansion, PBH formation and endothermic particle reactions. So TeV physics will modulate the PBH production rate and thus the mass distribution function function in the observable range.
Our toy model considered only geometry. Inhomogeneities and TeV physics, both LHC verified and BSM, would fill out the picture. An example is gg $\Leftrightarrow~\gamma \gamma$, where g is a gluon.
This provides a rich field for astroparticle physicists and microlensing astronomers to interact in. 
To bridge the gap, astronomers could speak loosely of a 1 TeV PBH, to identify its origin in the early Universe. 
\end{document}